\documentclass[aps,prl,twocolumn,superscriptaddress]{revtex4-1}
\usepackage{graphicx}
\usepackage{footnote}
\usepackage{epstopdf}
\usepackage{color}
\usepackage{amsmath}
\usepackage{multirow}

\begin{document}
\title{Measuring the Difference in Nuclear Charge Radius of Xe Isotopes by EUV Spectroscopy of Highly-Charged Na-like Ions}


\author{R. Silwal}
\email[]{rsilwal@g.clemson.edu}
\affiliation{Department of Physics and Astronomy, Clemson University, Clemson, SC, 29634}
\affiliation{National Institute of Standards and Technology (NIST), Gaithersburg, MD, 20899}

\author{A. Lapierre}
\affiliation{National Superconducting Cyclotron Laboratory, Michigan State University, East Lansing, MI, 48824}
\author{J.D. Gillaspy}
\affiliation{National Institute of Standards and Technology (NIST), Gaithersburg, MD, 20899}
\affiliation{ National Science Foundation, Arlington, VA, 22314}
\author{J.M. Dreiling}
\altaffiliation{Present address: Honeywell, Broomfield, CO, 80021}
\affiliation{National Institute of Standards and Technology (NIST), Gaithersburg, MD, 20899}

\author{S.A. Blundell}
\affiliation{University of Grenoble Alpes, CEA, CNRS, INAC-SyMMES, 38000 Grenoble, France}
\author{Dipti}
\affiliation{National Institute of Standards and Technology (NIST), Gaithersburg, MD, 20899}
\author{A.~Borovik~Jr}
\altaffiliation{Present address: I. Physikalisches Institut, Justus-Liebig-Universit{\"a}t Gie{\ss}en, 35392 Giessen, Germany}
\affiliation{National Institute of Standards and Technology (NIST), Gaithersburg, MD, 20899}

\author{G. Gwinner}
\affiliation{Department of Physics and Astronomy, University of Manitoba, Winnipeg, MB R3T 2N2, Canada}
\author{A.C.C. Villari}
\affiliation{National Superconducting Cyclotron Laboratory, Michigan State University, East Lansing, MI, 48824}
\author{Yu. Ralchenko}
\affiliation{National Institute of Standards and Technology (NIST), Gaithersburg, MD, 20899}
\author{E. Takacs}
\affiliation{Department of Physics and Astronomy, Clemson University, Clemson, SC, 29634}
\affiliation{National Institute of Standards and Technology (NIST), Gaithersburg, MD, 20899}


\date{\today}

\begin{abstract}
The difference in mean-square nuclear charge radius of xenon isotopes was measured utilizing a new method based on extreme ultraviolet spectroscopy of highly charged Na-like ions. The isotope shift of the Na-like D1 (3s $^{2}$S$_{1/2}$ - 3p $^2$P$_{1/2}$) transition between the $^{124}$Xe and $^{136}$Xe isotopes was experimentally determined using the electron beam ion trap facility at the National Institute of Standards and Technology. The mass shift and the field shift coefficients were calculated with enhanced precision by relativistic many-body perturbation theory and multi-configuration Dirac-Hartree-Fock method. The mean-square nuclear charge radius difference was found to be $\delta<r^2>^{136, 124}$ = 0.269(42) fm$^2$. Our result has smaller uncertainty than previous experimental results and agrees with literature values.
\end{abstract}


\maketitle

The charge radius and mass of the atomic nucleus ground state are amongst its most fundamental properties. Studies of nuclear charge radii are essential to understanding nuclear structure \cite{Hammen2018, Ruiz2016}. In particular, they have revealed unusual properties such as the large shape staggering in the neutron-deficient mercury isotope \cite{Bonn1972}, contributed to precision tests of the Standard Model \cite{Mane2011}, and enter in the determination of stellar element abundances \cite{Aoki2001}.

Only a few complementary techniques exist today for the determination of the absolute mean-square nuclear charge radius $<r^2>$, and its isotope variation $\delta<r^2>$. Muonic-atom spectroscopy \cite{Fricke1995} has been highly successful in the absolute measurement of $<r^2>$. Generally its accuracy is limited by large nuclear polarization effects as the muon orbit is comparable to the nuclear size. Electron scattering has also been widely used for the determination of the same quantity in heavy nuclei \cite{Hofstadter1956}, where the challenge is that the experimental cross-sections have to be analyzed beyond the first Born approximation to take into account the phase shift. Both methods require considerable amounts of target material and, with exception of recent efforts \cite{Tsukada2017}, are generally applied to stable nuclei.

X-ray spectroscopy of inner-shell K$\alpha$ lines and valence-electron optical isotope shifts allow for $\delta<r^2>$ measurements between isotopes \cite{Brockmeier1965}. The laser spectroscopy measurements of the latter technique offer utmost experimental precision and can be applied to long chains of stable and unstable isotopes \cite{Campbell2016}. The difficulty of this technique lies in the calculation of the electronic structure of many-electron atomic systems that often include electron correlation effects and can contribute to systematic offsets in the inferred $\delta<r^2>$ \cite{Libert2007}. Electron screening or correlation effects in heavy elements, such as bismuth or uranium, can be particularly difficult to calculate theoretically. These calculations are sometimes benchmarked by non-optical methods such as K$\alpha$ measurements \cite{Anatassov1992} or King plot analyses \cite{Cheal2012}.

In the search for new methods for the measurement of nuclear radii, particular charge states of highly ionized atoms have been considered due to their simpler electronic structure and higher sensitivity to the nuclear charge distribution. Their compressed electron cloud can produce large isotope shifts of energy levels. Relativistic normal and specific mass shifts have been explored through magnetic-dipole transitions of Be-like and B-like argon isotopes in the visible range \cite{Orts2006}, but the experimental precision was insufficient to probe the charge distribution. Precision X-ray spectroscopy \cite{Elliott1998} and dielectronic recombination measurements \cite{Schuch2005, Brandau2008} of isotope shifts in heavy, few-electron ions, have been used to determine $\delta<r^2>$ in a variety of nuclei because the electronic structure of these ions can be calculated with high accuracy. 

In this letter we introduce a new method based on accurate theoretical calculations for low-lying energy levels of Na-like ions. The simple 3s electronic configuration in these systems penetrates the Ne-like closed shell to probe the nucleus. Spectroscopic measurements of extreme ultraviolet (EUV) transitions are sensitive enough to determine nuclear charge distribution as previously discussed by Gillaspy {\em et al}. \cite{Gillaspy2013}. Atomic-structure calculations for these systems can reach accuracies higher than those for neutral atoms and singly charged ions used in optical isotope shift measurements.

Here, we present the first experiment using this technique to determine $\delta<r^2>$ of xenon isotopes by combining accurate theoretical calculations with precise measurement of the isotope shift in the frequency of the 3s $^{2}$S$_{1/2}$ - 3p $^{2}$P$_{1/2}$ (D1) transition in highly charged Na-like $^{136}$Xe$^{43+}$ and  $^{124}$Xe$^{43+}$ ions. Benchmarks of the quantity for this isotope pair are the previous optical isotope shift measurement that reported a value of 0.242(80) fm$^2$ \cite{Borchers1989}, the recommended value of 0.290(69) fm$^2$ that considers interconnected trends across the nuclear radii surface in a compilation by Angeli and Marinova \cite{Angeli2013}, as well as the detailed case-by-case analysis of Fricke and Heilig \cite{Fricke2004} yielding 0.324(57) fm$^2$. The new technique uses an electron beam ion trap (EBIT), which is similar to that previously employed for investigating unstable nuclei in Elliott {\em et al.} \cite{Elliott1998}.

The quantity determined experimentally is the $\delta \nu^{A, A'}$ isotope-dependent frequency shift, which has two components:
\begin{equation}
\delta \nu^{A, A'} = \delta \nu_{MS}^{A, A'} + \delta \nu_{FS}^{A, A'}.
\end{equation}
$\delta \nu_{MS}^{A, A'}$ is the mass shift due to the finite mass of the nucleus, and $\delta \nu_{FS}^{A, A'}$ is the field shift associated with the nuclear volume. It is notable that the field shift scales with the nuclear charge as approximately $Z^{8/3}$, and it dominates the mass shift in heavy systems. As an approximation, $\delta \nu_{FS}^{A, A'}$ can be considered to be proportional to the difference between the mean-square nuclear charge radii of the two isotopes:
\begin{equation}
\delta \nu_{FS}^{A, A'} = F \delta<r^2>^{A, A'},
\label{eq:eqFS}
\end{equation}
where $\delta<r^2>^{A, A'}$ is defined in Ref. \cite{Angeli2013}:
\begin{equation}
\delta<r^2>^{A, A'} = <r^2>^A - <r^2>^{A'}.
\end{equation}

Both the field shift coefficient,  $F$, and the mass shift, $\delta \nu_{MS}^{A, A'}$, can be obtained from highly accurate atomic-structure calculations, allowing for the experimental determination of $\delta<r^2>^{A, A'}$ from the measured $\delta \nu^{A, A'}$ shift.

In this experiment, EUV spectra  were collected from Na-like $^{136}$Xe$^{43+}$ and $^{124}$Xe$^{43+}$ ions produced in the EBIT at the National Institute of Standards and Technology (NIST). Details of the measurements of EUV emission from xenon ions are similar to that in previous experiments in this wavelength region \cite{Gillaspy2013}.

Briefly, over the course of the experimental campaign, isotopically pure $^{136}$Xe and $^{124}$Xe neutral gases were alternately injected into the
EBIT for approximately one-hour periods at a time. For each gas injection, a series of spectra were collected for five minutes each, using a
liquid-nitrogen-cooled EUV charge-coupled-device (CCD) camera attached to a flat-field EUV grating spectrometer \cite{Blagojevic2005, Silwal2017}.

\begin{figure}[t!]
\includegraphics[width=8.6 cm]{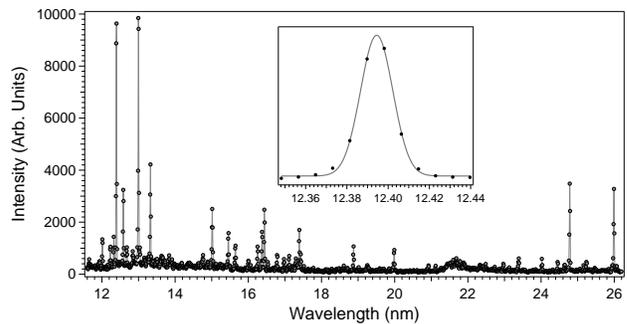}
\caption{Sample of a five-minute EUV spectrum of $^{136}$Xe with spectral lines from different charge states. The inset shows a blowup of the Na-like D1 transition in the first order of diffraction, along with a Gaussian fit.}
\label{fig:XeD1}
\end{figure}

The EBIT was operated at 6.0 keV electron beam energy and 150 mA electron beam current to optimize the production of the Na-like Xe charge state. Na-like D1 transition was selected for the determination of $\delta<r^2>^{136, 124}$ because of the accuracy of the calculations for this line and because it is cleanly separated unlike the 3s $^{2}$S$_{1/2}$ - 3p $^2$P$_{3/2}$ (D2) line that was effected by a blend \cite{Gillaspy2010}. Figure \ref{fig:XeD1} shows the full spectral range detected by the CCD camera, including emission from Na-like Xe and nearby charge states.

Absolute wavelength calibration \cite{Silwal2017} was carried out using well-known transitions in different charge states of Ne, Xe, and Ar \cite{Kramida2017} collected several times during the data-taking epoch. The first derivative of the absolute calibration function provided the dispersion function to convert the measured spatial shift on the CCD chip to a wavelength shift. The isotope shift of the D1 transition in this experiment was well within the uncertainty of the absolute wavelength value of 12.3935(9) nm.

The experiment was a multi-day acquisition effort during which long term thermal and electronic systematic drifts could be expected. A large number of photons were required to achieve the necessary statistical uncertainty due to the less than 10$^{-4}$ nm anticipated shift between the D1 lines of the two xenon isotopes. To account for the long-term variation of the D1 line position, we determined the centroid positions in each individual 300 s spectrum and created a time-ordered sequence of the channel positions that included both isotopes.

The full sequence of 428 values was fitted to a set of third order polynomial functions that described the long-term variation of the absolute position of the spectral line. Different sets of coefficients were allowed for datasets in between liquid nitrogen refills of the CCD camera, at which times more thermal variations were expected. The polynomial functions for the two isotopes were kept to be the same except for an overall constant free parameter representing the isotope shift.

\begin{figure}[htb!]
\begin{minipage}[t] {\linewidth}
\includegraphics[width=8.6 cm]{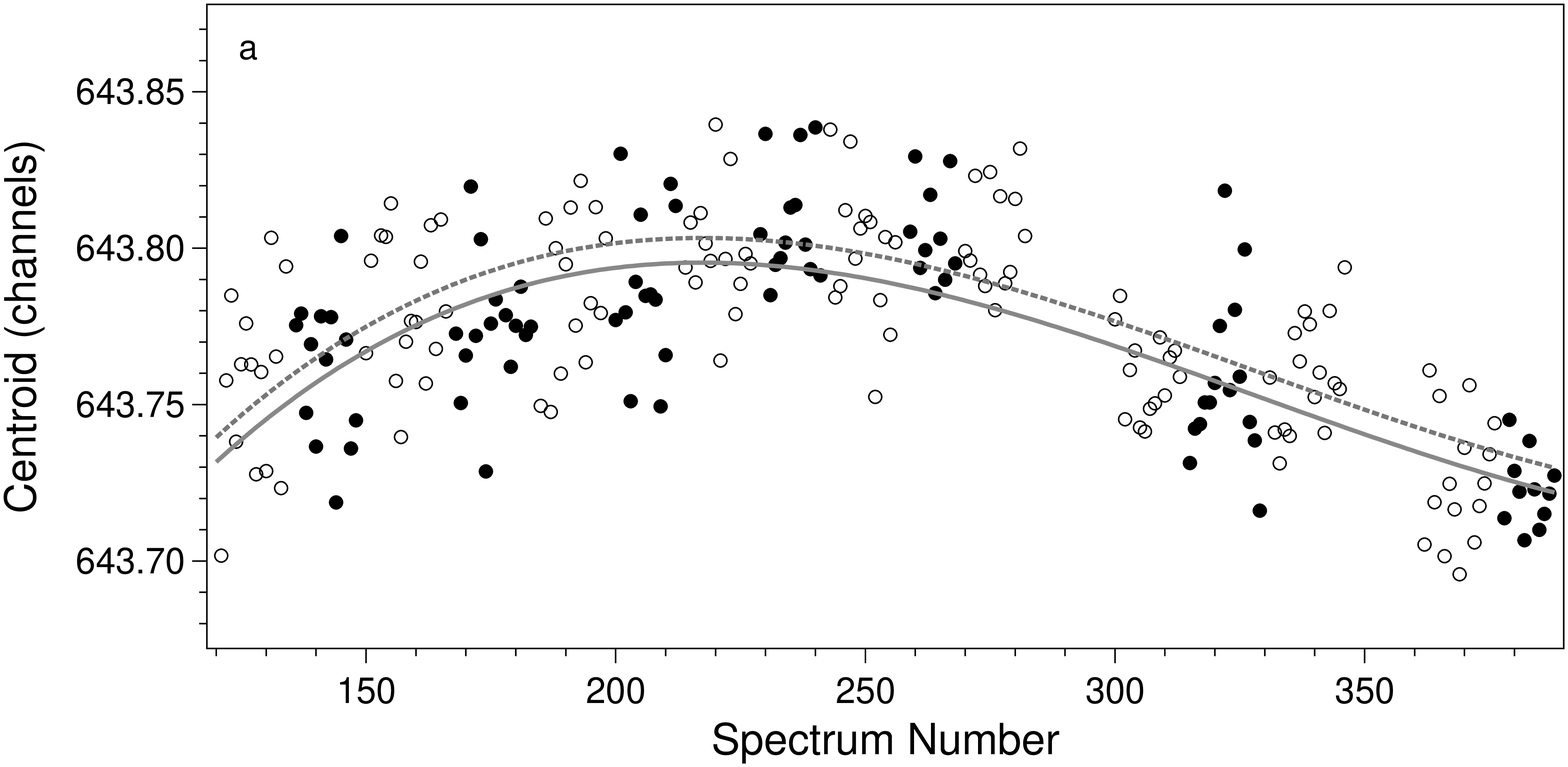}
\end{minipage}

\hspace*{\fill} 

\begin{minipage}[t] {\linewidth}
\includegraphics[width=8.2 cm]{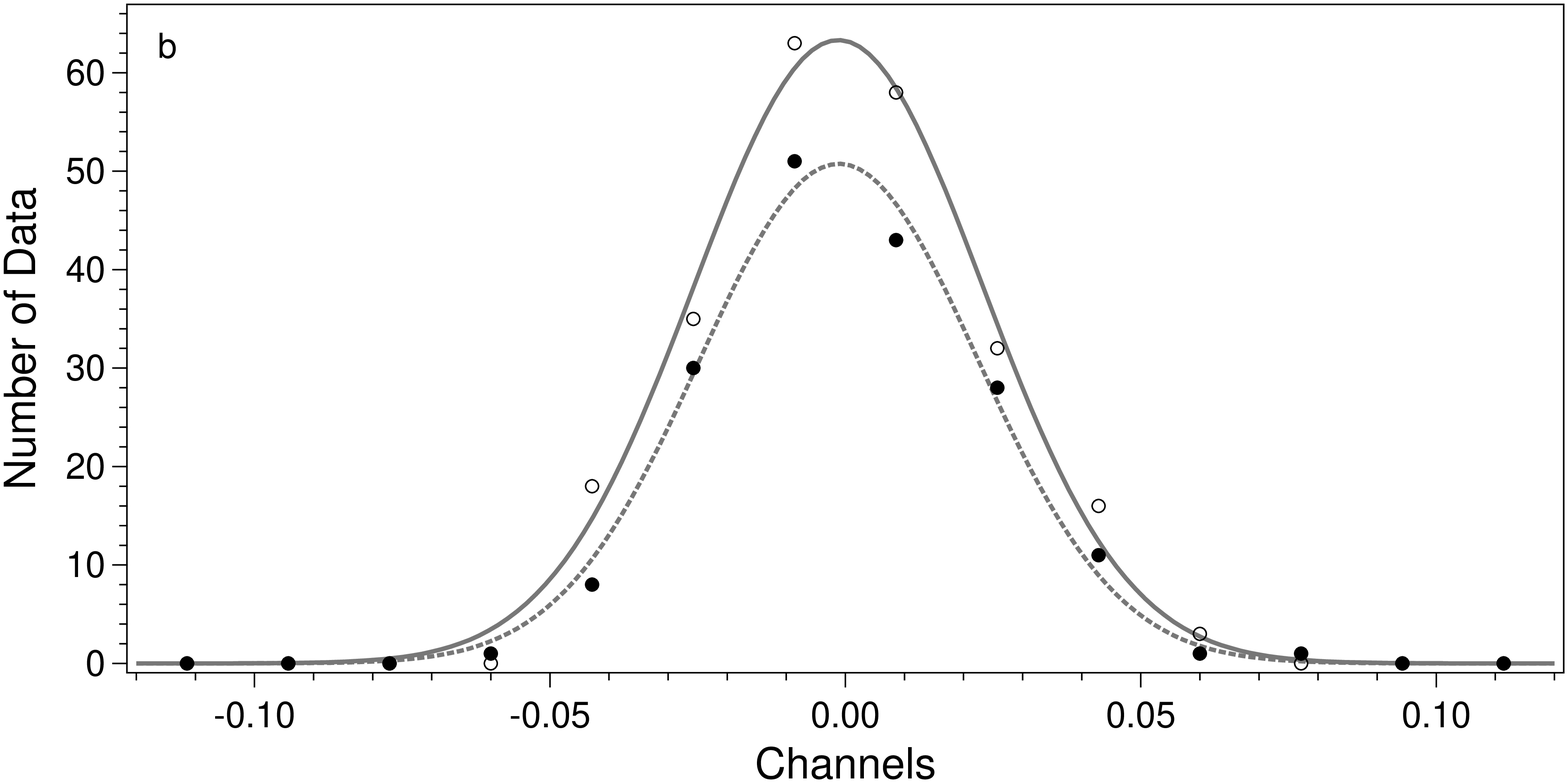}
\end{minipage}
\caption{a) Partial series of the centroid positions of the Na-like D1 transition in first order for the $^{124}$Xe (open circles) and $^{136}$Xe (full circles) isotopes and their fitted values described in the text. b) Statistical distributions of the residuals $^{124}$Xe (open circles) and $^{136}$Xe isotopes  (full circles) fitted with pure Gaussian functions.}
\label{fig:scatter}
\end{figure}

Figure \ref{fig:scatter}a shows a partial series of the centroid values with the alternating isotope sequences, the polynomial fit, and the constant isotope shift. Residuals were binned for each isotope individually, providing statistical distributions, which were fit with pure Gaussian functions shown in Fig. \ref{fig:scatter}b. The agreement between the centroids of the Gaussian functions is more than an order of magnitude better than the uncertainty of the shift, giving us confidence in the evaluation procedure.

To verify the consistency of the experimental results, a series of systematic tests of different polynomial orders for the systematic drift, the number of channels in calculating the centroid positions, and the residual distribution bin sizes were performed. The overall experimental uncertainty of the determined 6.5 $\times$ 10$^{-5}$ nm wavelength shift between the two isotopes was 2.1 $\times$ 10$^{-5}$ nm. It was dominated by the 2.0 $\times$ 10$^{-5}$ nm statistical uncertainty associated with the spatial shift determination and by the 0.4 $\times$ 10$^{-5}$ nm uncertainty due to the dispersion function which also include a systematic component. The experimental analysis was concluded by converting the wavelength shift into a frequency shift $\delta\nu^{A, A'}$ to determine $\delta <r^2>^{A, A'}$ based on the following evaluation procedure.

The $\delta\nu_{MS}^{A, A'}$ mass shift (MS) and $\delta\nu_{FS}^{A, A'}$ field shift (FS) for the Na-like D1 (3s $^{2}$S$_{1/2}$ - 3p $^2$P$_{1/2}$) transition were calculated using two different theoretical methods: the relativistic many-body perturbation theory (RMBPT) \cite{Johnson1988A, Johnson1988B} and the multi-configuration Dirac-Hartree-Fock (MCDHF) theory of the GRASP2K code \cite{Jonsson2013}.

RMBPT was performed up to third order including both the Coulomb and Breit interactions in each order. A relativistic configuration interaction (RCI) module was used in the GRASP2K code to consider the Breit interaction perturbatively as well as leading quantum electrodynamics (QED) contributions (vacuum polarization and self-energy corrections). The two theories were in an overall excellent agreement. Table \ref{table1} shows the experimental and calculated isotope shifts along with calculations from the large-scale configuration-interaction Dirac-Fock (CIDF) method by Tupitsyn {\em et al.} \cite{Tupitsyn2003} solving the Dirac-Coulomb-Breit equation.

To obtain the field shift from the experimental $\delta\nu^{A, A'}$ frequency shift, the mass shift was accounted for through theoretical calculations. In RMBPT theory, the relativistic nuclear recoil corrections were calculated up to order $(Z\alpha)^2$ beyond the nonrelativistic mass shift by using the Palmer operator \cite{Palmer1987}. This operator gives the one- and two-body nuclear-recoil terms in the relativistic Hamiltonian corresponding to the $\delta\nu_{NMS}^{A, A'}$ normal mass shift (NMS) and $\delta\nu_{SMS}^{A, A'}$ specific mass shift (SMS), respectively. The nonrelativistic nuclear-recoil effect is itself of the order of $(Z\alpha)^2$; therefore, the leading relativistic correction considered here is of the order of $(Z\alpha)^4$. The mass shift in each order of RMBPT was determined by the difference of calculations with and without nuclear recoil, and the difference was tested for numerical significance. This provided the NMS and the SMS coefficients, R and S, respectively, defined such that the $\delta\nu_{NMS}^{A, A'}$ and $\delta\nu_{SMS}^{A, A'}$ frequency shifts for nuclear masses M$_A$ and M$_{A'}$ are given by:
\begin{equation}
\delta\nu_{NMS}^{A,A'} = R(1/M_A - 1/M_{A'}),
\end{equation}
and
\begin{equation}
\delta \nu_{SMS}^{A,A'} = S(1/M_A - 1/M_{A'}),
\end{equation}
where
\begin{equation}
\delta\nu_{MS}^{A, A'} = \delta\nu_{NMS}^{A, A'} + \delta\nu_{SMS}^{A, A'}.
\end{equation}

Third-order RMBPT contributions to the mass shift coefficients for the D1 transition were on the order 0.1 \% or less of the total mass shift. The dominant theoretical uncertainty in the mass shift is in the omitted higher-order relativistic terms starting at the order of $(Z\alpha)^5$ \cite{Erickson1965}. We assumed 5 \% of the total mass shift throughout.

For the field shift coefficient $F$ in RMBPT, the transition frequency $\nu_{A}$ was calculated for isotopes $A$ = 136 and $A'$= 124, omitting the NMS and SMS contibutions and assuming a two-parameter (half density radius and surface thickness) Fermi model nuclear charge distributions. The difference $\delta \nu_{FS}^{A, A'}$ =  $\nu_{A}$ - $\nu_{A'}$ was taken in each order of RMBPT. The RMBPT calculations for the field shift were found to converge rapidly, with the third-order contributions to the D1 transition less than 0.1 \% of the total $\delta\nu_{FS}^{A, A'}$. The field-shift coefficient was obtained as $F$ = $\delta\nu_{FS}^{A, A'}$/$\delta<r^2>^{A, A'}$, where $\delta<r^2>^{A, A'}$ is the change in mean-square radius.

The dominant theoretical uncertainty of $F$ comes from the unknown nuclear charge distributions. By calculating $F$ for several pairs of nuclear parameters, it was found that $F$ fluctuated on the 2 \% level, which we take to be the associated uncertainty.

In the MCDHF approach, the atomic state function was expanded in the configuration state functions of the same parity, total angular momentum ($J$), and its projection ($M_J$). The reference configurations were 1s$^2$2s$^2$2p$^6$3s and 1s$^2$2s$^2$2p$^6$3p for the ground and excited states, respectively, and the single configuration Dirac-Fock state functions were calculated for the $^{136}$Xe isotope. The Breit interactions and the leading QED effects up to n= 5 were included during RCI calculations. Self-energy and vacuum polarization QED corrections were estimated phenomenologically and were found to enter at the 0.1 \% level.

The relativistic isotope shift (RIS3) module \cite{Naze2013} was used to calculate the mass shift from the wave functions. Similar to RMBPT, GRASP2K also includes nuclear recoil corrections of the order of $(Z\alpha)^4$ for mass shift calculations. The field shift was calculated explicitly from the difference between transition energies that were obtained by solving the MCDHF and Breit equations for isotopes $A$ and $A'$ separately.

For a transition involving a valence $s$ electron, Eq. \ref{eq:eqFS} can be more accurately replaced by
\begin{equation}
 \delta \nu_{FS}^{A, A'} = F \lambda_{Seltzer}^{A, A'},
\end{equation}
where $\lambda_{Seltzer}^{A, A'}$ is the Seltzer moment of the nucleus \cite{Blundell1987}:
\begin{eqnarray*}
\lambda_{Seltzer}^{A, A'} = \delta<r^2>^{A, A'}+S_4 \delta<r^4>^{A, A'} + S_6 \delta<r^6>^{A, A'} \\
= [1 + S_{HO}/\delta<r^2>^{A, A'}]\delta<r^2>^{A, A'},
\end{eqnarray*}
with $S_{HO}$ representing the higher nuclear moment terms. The values of the $S_4$ and $S_6$ coefficients \cite{Blundell1987} suggest that these contributions to $\delta \nu_{FS}^{A, A'}$ were 4 \% in our case.

Similar conclusion was drawn from GRASP2K calculations of the field shift \cite{Li2012}, which used the probability density of the electron wave function at the origin effectively selecting the term $\delta<r^2>^{A, A'}$ in $\lambda_{Seltzer}^{A, A'}$. The field shift obtained this way was 4 \% larger than the result implicitly containing higher-order nuclear moments.

\begin{table}[htb!] 
\caption{Measured and calculated wavelength values of the isotope shift along with their uncertainties (in units of fm) for the Na-like D1 transition 3s $^{2}$S$_{1/2}$ - 3p $^2$P$_{1/2}$ for the isotope pair $^{136}$Xe -- $^{124}$Xe. The field shift was calculated using the evaluated value of 0.290 fm$^2$ for $\delta <r^2>^{136,124}$ by \cite{Angeli2013}.}
\label{table1}
\begin{ruledtabular}
\begin{tabular}{cccccccc}
             & \multicolumn{5}{c}{\textbf{Theory}}                                                                                                 & \multicolumn{2}{c}{\multirow{2}{*}{\textbf{Experiment}}} \\
             & \multicolumn{2}{c}{RMBPT}                  & \multicolumn{2}{c}{GRASP2K}                & CIDF \cite{Tupitsyn2003} & \multicolumn{2}{c}{}                                     \\ \hline
Coefficients & $\delta \lambda$ & $\Delta \delta \lambda$ & $\delta \lambda$ & $\Delta \delta \lambda$ & $\delta \lambda$                          & $\delta \lambda$           & $\Delta \delta \lambda$     \\ \hline
NMS          & -4.8             & 0.2                     & -4.8             & 0.2                     & -4.8                                      &     			  &       \\
SMS          & -62.2            & 3.4                     & -62.3            & 3.4                     & -62.7                                     &                            &                             \\
Total MS     & -67.0            & 3.4                     & -67.1            & 3.4                     & -67.5                                     &                             &                             \\
FS           & 143.0            & 2.8                     & 142.0            & 2.8                     & 143.0                                     &                                &                            \\
Total        & 76.1             & 4.4                     & 75.3             & 4.4                     & 75.8                                      &     65.5                 &               20.6 \\
\end{tabular}
\end{ruledtabular}
\end{table}

Using the calculated values of the mass shift and $F$, along with the experimentally obtained frequency shift $\delta\nu$, the difference between the mean-square nuclear charge radii of $^{136}$Xe and $^{124}$Xe was determined. The value obtained using the RMBPT theoretical method was 0.268(42) fm$^2$, and using the GRASP2K results it was 0.270(42) fm$^2$. Our reported value is their average of $\delta<r^2>^{136,124}$ = 0.269(42) fm$^2$.

The overall one sigma uncertainty $\Delta$ $\delta<r^2>^{136, 124}$ includes uncertainties from the experimental shift, the mass shift correction, the field shift calculation, and the higher order nuclear moments. We note that the discussion of hyperfine effects is beyond the scope of this paper, as both nuclei have zero nuclear spin.

Our 16 \% relative total uncertainty of $\delta<r^2>^{136, 124}$ is mainly due to the experimental uncertainty dominated by the counting statistics. The theoretical uncertainty amounts to about 3 \% including the mass-shift. The different quantities that contribute to the evaluation of $\delta<r^2>^{136, 124}$ together with their measured or estimated uncertainties are listed in Table \ref{table1}.

\begin{figure}[htb!]
\includegraphics[width=8.6 cm]{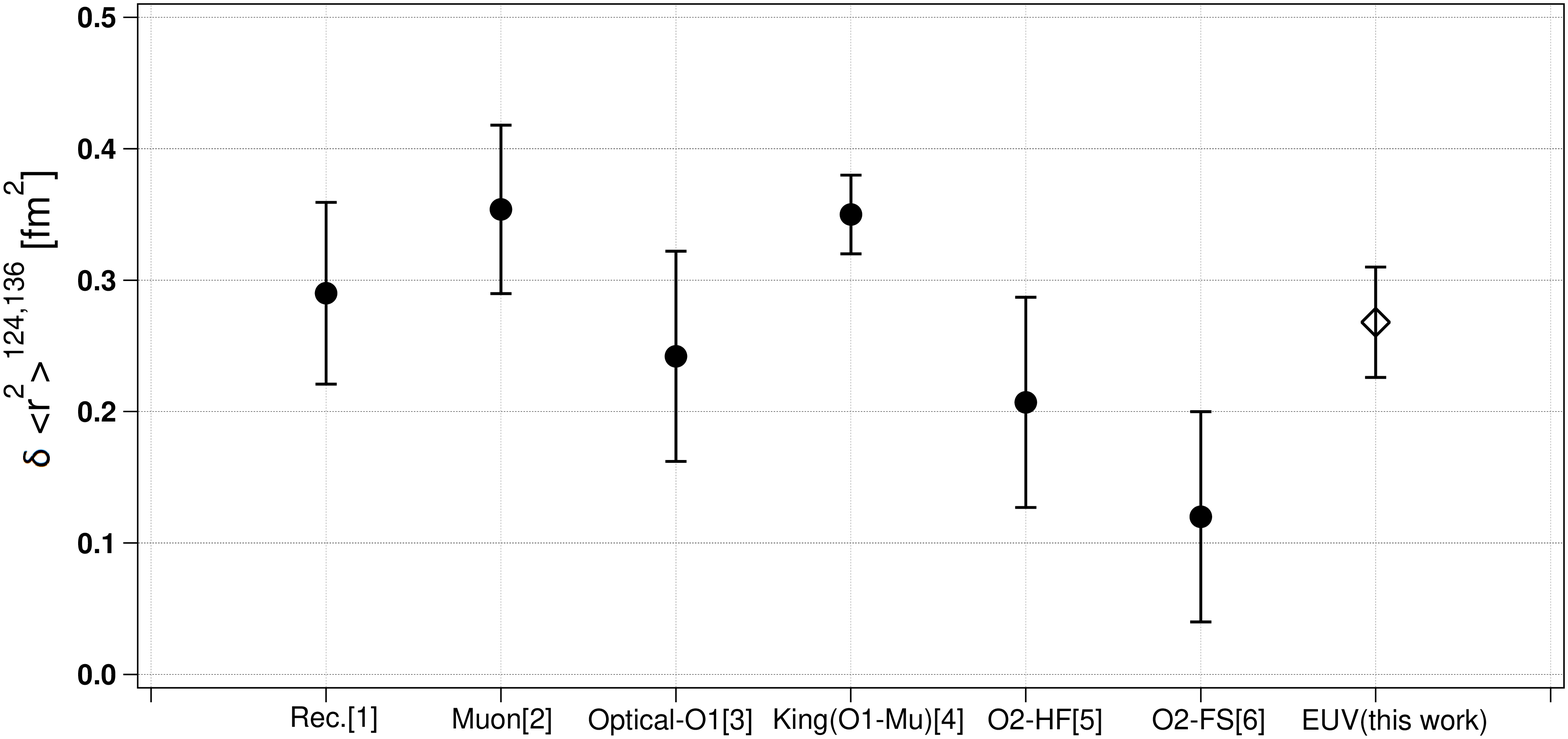}
\caption{$\delta<r^2>^{136,124}$ change in mean-square nuclear charge radius between $^{136}$Xe and $^{124}$Xe measured in this work through Na-like D1 EUV spectroscopy (open diamond) compared with previous measurements and analyses (circles). Rec.: recommended value by \cite{Angeli2013}; Muon: muonic-atom spectroscopy \cite{Fricke1995}; Optical-O1: optical shift by laser-spectroscopy \cite{Borchers1989}; King (O1-Mu): King plot analysis combining the optical measurements with the muonic-atom results \cite{Libert2007}; O2-HF and O2-FS: optical (interferometer) shift based on the Hartree-Fock method and Fermi-Segre calculations \cite{Fischer1974}.}
\label{fig:comparison}
\end{figure}

Figure \ref{fig:comparison} presents the result of the current experiment compared to five previous values using various techniques. Our result agrees within its uncertainty with the values of 0.290(69) fm$^2$ evaluated by \cite{Angeli2013} and 0.242(80) fm$^2$ obtained from the optical (laser spectroscopy) isotope shift measurement by Borchers {\em et al.} \cite{Borchers1989}. Their combined 0.080 fm$^2$ uncertainty includes 0.005 fm$^2$ experimental and an order of magnitude larger 0.080 fm$^2$ theoretical uncertainties due to the lack of precise theoretical calculations for the neutral system. 

Libert {\em et al.} \cite{Libert2007} performed nuclear-structure calculations including dynamical deformation of the evolution of the mean-square charge radius of xenon over a long isotope chain. Their results were compared with experimental charge radii deduced by combining Borchers {\em et al.} \cite{Borchers1989} optical isotope shifts with an $F$ value obtained from semi-empirical atomic-structure calculations and from a King-plot analysis including muonic atom results. It was found that the model is in excellent agreement with the charge radii calculated with the semi-empirical $F$ value and disagrees with the predictions of the King plot. Libert {\em et al.} \cite{Libert2007} argued that the disagreement is because the muonic charge radii were measured near the magic neutron number N=82, where the charge radius change and the nuclear polarization corrections are small. Indeed, the latter was calculated by assuming spherically shaped nuclei, whereas deformation is known to exist for the lightest stable nuclei, therefore $F$ values obtained from those stable nuclei might not be accurate for nuclei far from stability. 

Our result is outside the error bar of the result of 0.350(30) fm$^2$ predicted by the King-plot analysis, and it is within the uncertainty of the optical result obtained with the semi-empirical $F$ value \cite{Libert2007}. This finding is an experimental support for the importance of the dynamical deformation along the xenon isotopic chain especially in the analysis of muonic data.

In conclusion, EUV spectroscopy of Na-like ions is a viable independent method for accurate nuclear size measurements for heavy nuclei. The current theoretical and experimental systematic effects are on the order of a few percent for medium heavy systems. The overall uncertainty can approach this level by increasing counting statistics, allowing the systematic study of subtle changes of the nuclear radius along sequences. Na-like ions can be produced in large abundance in EBIT devices, offering the possibility of conducting measurements on radioactive isotopes at existing rare-isotope beam facilities like the NSCL \cite{Lapierre2018} and TITAN/TRIUMF \cite{Lapierre2010}. Present and next-generation facilities where this method could be implemented are CARIBU \cite{Ostroumov2015}, ISOLDE \cite{Wenander2010}, FRIB \cite{Glasmacher2017}, RAON/RISP \cite{Kim2015}, and MATS/FAIR\cite{Rodriguez2010}.

\begin{acknowledgments}
This work was partially funded by the NIST Grant Award \# 70NANB16H204 of the Measurement Science and Engineering (MSE) Research Grant Programs. AL
and ACCV acknowledges support by the National Science Foundation under Grant No. PHY-1565546. JMD acknowledges funding from the National Research
Council Postdoctoral Fellowship at NIST. GG acknowledges support by NSERC (Canada). ABJr acknowledges support of the US Department of Commerce
and NIST under the program \# G-3-00334. We would like to thank David Takacs for his help with the initial data analysis.
\end{acknowledgments}


\end{document}